\documentstyle[aps,multicol,psfig]{revtex}

\begin{document}

\draft

\title{Phase Transitions in Rotating Neutron Stars}

\author{Henning Heiselberg and Morten Hjorth-Jensen}

\address{NORDITA, Blegdamsvej 17, DK-2100 Copenhagen \O, Denmark}

\maketitle

\begin{abstract}
 
As rotating neutron stars slow down, the pressure and the density in
the core region increase due to the decreasing centrifugal
forces and phase transitions may occur in the center. We extract the analytic
behavior near the critical angular velocity $\Omega_0$, where the
phase transitions occur in the center of a neutron star, and calculate
the moment of inertia, angular velocity, rate of slow down, braking
index, etc.  For a first order phase transition these quantities
have a characteristic behavior, e.g., the braking index diverges as
$\sim(\Omega_0-\Omega)^{-1/2}$.  Observational consequences for first,
second and other phase transitions are discussed.

\end{abstract}


\begin{multicols}{2}

The physical state of matter in the interiors of neutron stars at
densities above a few times normal nuclear matter densities is
essentially unknown. Interesting phase transitions in nuclear matter 
to quark matter \cite{star_properties},
mixed phases of quark and nuclear matter \cite{Glendenning,HPS}, kaon
\cite{Kaplan} or pion condensates \cite{pion,vijay}, 
neutron and proton superfluidity \cite{oeystein},
hyperonic matter \cite{star_properties,Glendenning},
crystalline nuclear matter \cite{pion}, magnetized matter, etc., have
been considered.  Recently, Glendenning et al. \cite{GPW} have
considered the interesting question of rapidly rotating neutron stars
and what happens as they slow down when the decreasing centrifugal
force leads to increasing core pressures.  They find that a drastic
softening of the equation of state, e.g., by a phase transition to
quark matter can lead to a sudden contraction of the neutron star at
a critical angular velocity and shows up in a backbending moment of
inertia as function of frequency.  Here we consider another
interesting phenomenon namely how the star and in particular its moment
of inertia behaves near the critical angular velocity where the core
pressure just exceeds that needed to make a phase transition. We
calculate the moment of inertia, angular velocities, braking
index, etc.\ near the critical angular velocity and discuss
observational consequences for first and second order phase
transitions.

The general relativistic equations for slowly rotating stars were 
described by Hartle \cite{Hartle}. 
We shall also make the standard approximation of slowly
rotating stars, i.e., the rotational angular velocity is
$\Omega^2 \ll GM/R^3$.
For neutron stars with mass $M=1.4M_\odot$ and
radius $R\sim 10$ km their period should thus be larger than a few
milliseconds, a fact which applies to all measured pulsars insofar.
Hartle's equations are quite elaborate to solve
as they consist of six coupled differential equations as compared to
the single Tolman-Oppenheimer-Volkoff 
equation \cite{TOV} in the non-rotating case.
In order however to be able to analytically
extract the qualitative behavior near the critical angular velocity
$\Omega_0$, where a
phase transition occurs in the center, we will first solve the
Newtonian equations for a simple equation of state.
This will allow us to make general predictions on properties
of rotating neutrons stars when phase transitions occur in the interior
of a star.
The corrections from general relativity are typically of order
$GM/R\simeq$10\% for neutron stars of mass $M\simeq 1.4M_\odot$.
The extracted analytical properties of a rotating star 
are then  checked below by actually solving 
Hartle's equations numerically for a realistic equation of state.
 
The simple Newtonian equation of
motion expresses the balance between the pressure gradient and 
the gravitational and centrifugal forces
\begin{equation}
   \nabla P = -\rho\left( \nabla V +
        {\bf \Omega}\times{\bf \Omega}\times{\bf r} \right),
   \label{NOV}
\end{equation}
Here, $V({\bf r})$ is the gravitational potential for the deformed
star and $\rho$ the energy ($\sim$mass) density.  We assume that friction in the 
(nonsuperfluid) matter insures
that the star is uniformly rotating.  Since cold neutron stars are
barotropes, i.e., the pressure is a function of density,
the pressure, density and effective gravitational potential, 
$\Phi=V-\frac{1}{2}({\bf \Omega}\times{\bf r})^2$, are all
constants on the {\it same} isobaric surfaces
for a uniformly rotating star \cite{Hartle}.
We denote these surfaces by the effective radius, $a$, and
for slowly rotating stars it is related to the distance $r$ from the
center and the polar angle $\theta$ from the rotation axis along
${\bf\Omega}$ by \cite{Hartle}
\begin{equation}
   r(a,\theta) = a\left[ 1-\epsilon(a) P_2(\cos\theta) \right],
   \label{ra}
\end{equation}
where $P_2(\cos\theta)$ is the 2nd Legendre polynomial and
$\epsilon(a)$ is the deformation of the star from spherical symmetry.

Inserting Eq.\ (\ref{ra}) in Eq.\ (\ref{NOV}) 
one obtains for small deformations 
\cite{Hartle} the $l=0$ Newtonian hydrostatic equation
\begin{equation}
  \frac{1}{\rho}\frac{dP}{da} = -G\frac{m(a)}{a^2} + \frac{2}{3}\Omega^2 a,
  \label{Pa} 
\end{equation}
where $m(a)=4\pi\int^a_0\rho(a')a'^2da'$ is the mass
contained inside the mean radius $a$.
The factor 2/3 in the centrifugal force arises because it only acts in
two of the three directions.
The equation $(l=2)$ for the deformation $\epsilon(a)$
is given in, e.g., Ref.\ \cite{Tassoul}. 
The deformation generally increases
with decreasing density, i.e., the star is more deformed in its outer layers.

In order to discuss the qualitative behavior near critical angular
velocities we first consider a simple EoS with phase transitions for which
Eq.\ (\ref{Pa}) can be solved analytically namely that of
two incompressible fluids with a first order phase transition
between energy density
$\rho_1$ and $\rho_2$ ($\rho_1<\rho_2$) 
coexisting at a pressure $P_0$.
The mass function $m(a)$ is very simple 
in the Newtonian limit and the boundary condition $m(a)=M$
relates the star radius $R$ to the radius of the dense core, $R_0$, as
\begin{equation}
    R = \left(\bar{R}^3-(\frac{\rho_2}{\rho_1}-1)R_0^3\right)^{1/3}, 
\end{equation}
where $\bar{R}=(3M/4\pi\rho_1)^{1/3}$ is the star radius
in the absence of a dense core.
Solving Eq.\ (\ref{Pa}) gives the pressure
\begin{eqnarray}
    P(a)&=&P_0+\frac{1}{2}(R_0^2-a^2)\rho_1(\frac{4\pi}{3}G\rho_1
     -\frac{2}{3}\Omega^2)\nonumber \\ 
    && +\frac{4\pi}{3}GR_0^2(\rho_2-\rho_1)\rho_1(1-\frac{R_0}{a}),
     \label{Pai2}
\end{eqnarray}
for $R_0\le a\le R$.
The boundary condition at the surface $P(R)=0$ in
Eq.\  (\ref{Pai2}) gives 
\begin{eqnarray}
  \omega^2 & \equiv &\frac{\Omega^2}{2\pi G\rho_1}
  = 1-2\left(\frac{3}{4\pi}\frac{P_0}{G\rho_1^2R^2} \right. \nonumber \\
    & & \left. +(\frac{\rho_2}{\rho_1}-1)\frac{R_0^2}{R^2}(1-R_0/R) \right)
               (1-R_0^2/R^2)^{-1}. \label{om}
\end{eqnarray}
The phase transition occurs right at the center when $R_0=0$ corresponding to
the {\it critical angular velocity} $\Omega_0=\omega_0\sqrt{2\pi G\rho_1}$
where
\begin{equation}
   \omega_0^2= 1-2\frac{P_0\bar{R}}{GM} 
   \, .\label{o0}
\end{equation}
Generally, for any EoS the critical angular velocity depends on $P_0$, $M$, 
and $\rho_1$ but not on $\rho_2$.

For angular velocities just below $\omega_0$ very little of 
the high density phase exists and $R_0\ll R$. Expanding (\ref{om}) we obtain
\begin{equation}
  \frac{R_0}{\bar{R}} \simeq \sqrt{\frac{\omega_0^2-\omega^2}
             {2\rho_2/\rho_1-1-\omega_0^2}}. 
  \label{R0}
\end{equation}
For $\omega\ge\omega_0$ the dense phase disappears and $R_0=0$.

The corresponding moment of inertia is for $R_0\ll R$  
\begin{eqnarray}
   I &=& \frac{4\pi}{5} \left( \rho_2R_0^5+\rho_1(R^5-R_0^5)\right)
     (1+\frac{2}{5}\epsilon) \nonumber \\
     &\simeq& \frac{2}{5}M\bar{R}^2 
    \left( 1-\frac{5}{3}(\frac{\rho_2}{\rho_1}-1)\frac{R_0^3}{\bar{R}^3}\right)
      (1+\frac{1}{2}\omega^2),
    \label{I}
\end{eqnarray}
where we used that the deformation is $\epsilon=(5/4)\omega^2$ in the low
density phase \cite{Tassoul}. 
However, for the qualitative behavior near $\Omega_0$ only the contraction
of the star radius $R$ with the appearance of the dense core $R_0$
is important whereas the deformations can be ignored.
The contraction is responsible for the term in
the moment of inertia and is proportional to
$R_0^3\propto (\omega_0^2-\omega)^{3/2}$ near the critical angular
velocity. Consequently,  the derivative $dI/d\omega^2$ displays the same
non-analytic square root dependence as $R_0$ (see Eq.\ (\ref{R0})). 

Let us subsequently 
consider a more realistic EoS for dense nuclear matter at high
densities such as the Bethe-Johnson EoS \cite{BJ}. At
high densities it can be approximated by a polytropic relation between
the pressure and energy density:
$P=K_1\rho^{2.54}$, where 
$K_1=0.021\rho_0^{-1.54}$ and $\rho_0=m_n 0.15$fm$^{-3}$
is normal nuclear matter mass density. 
As we are only interested in the dense core we will for simplicity employ 
this Bethe-Johnson polytrope (BJP) EoS.
The central density of a non-rotating
1.4$M_\odot$
mass neutron star with the BJP is $\sim 3.4\rho_0$.
Furthermore, we assume that a first order phase
transition occurs at density $\rho_1=3.2\rho_0$ to a high density phase
of density $\rho_2=4\rho_0$ with a similar polytropic 
EoS $P=K_2\rho^{2.54}$. From the Maxwell
construction the pressure is the same at the interface, $P_0$,
which determines $K_2=K_1(\rho_1/\rho_2)^{2.54}$. 
We now generalize Eq.\   (\ref{Pa}) by including 
effects of general relativity. From Einstein's field equations for the metric
we obtain from the $l=0$ part 
\begin{equation}
  \frac{1}{\rho+P}\frac{dP}{da} = - G\frac{m+4\pi a^3P}{a^2(1-2Gm/a)} 
                    + \frac{2}{3}\Omega^2 a , 
  \label{PaGR}
\end{equation}
where $m(a)=4\pi\int_0^a\rho(a')a'^2da'$.
In the centrifugal force term
we have ignored frame dragging and other corrections of order
$\Omega^2GM/R\sim 0.1\Omega^2$ for simplicity and 
since they have only minor effects in our case.
By expanding the pressure, mass function and gravitational potential
in the difference between the rotating and non-rotating case, 
Eq.\  (\ref{PaGR}) reduces to the $l=0$ part of
Hartle's equations (cf. Eq.\  (100) in \cite{Hartle}.)
Note also that Hartle's full 
equations cannot be used in our case because the first order
phase transition causes discontinuities in densities so that changes
are not small locally. This shows up, for example, in the divergent
thermodynamic derivate $d\rho/dP$.

\begin{figure}
\centerline{
\psfig{figure=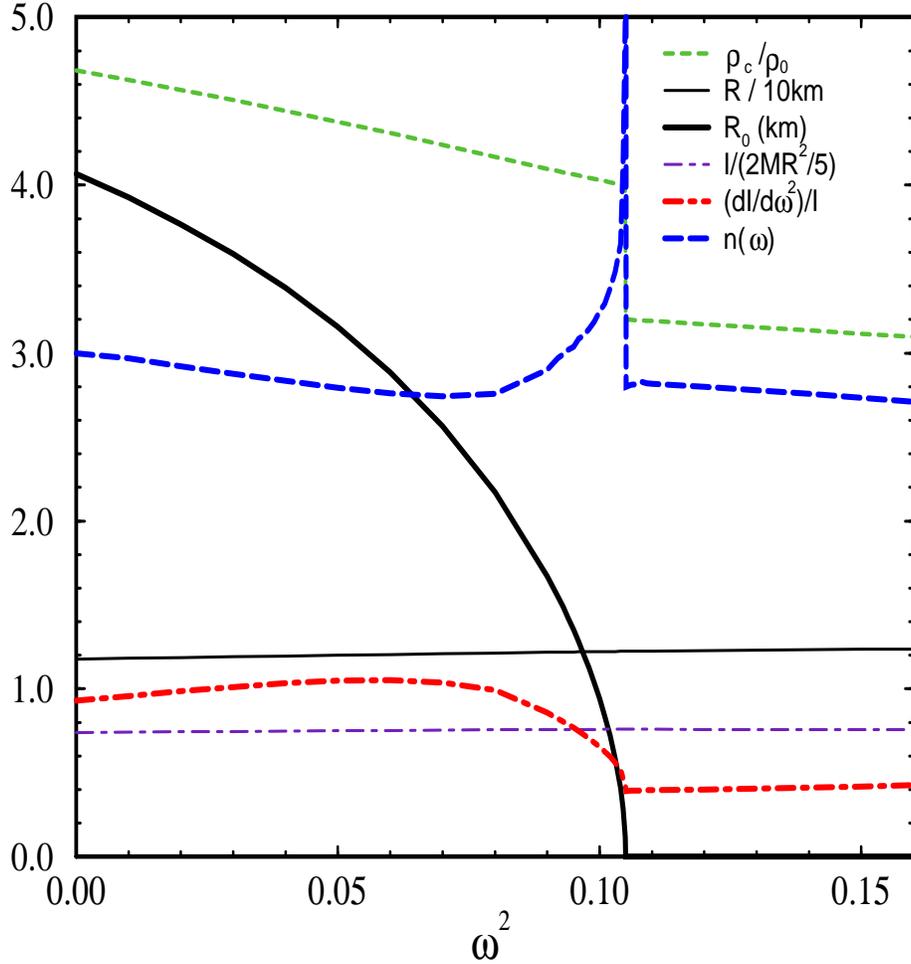,width=15cm,height=15cm,angle=-90}
}
\caption{ 
Central density (in units of $\rho_0$), radii of the neutron star $R$ and 
its dense core $R_0$, moment of inertia, its derivative
$I'/I=dI/d\omega^2/I$ and the braking index
are shown as function of the scaled angular velocity
$\omega^2=\Omega^2/(2\pi G\rho_1)$.
The rotating neutron star has mass $1.4M_\odot$ and a Bethe-Johnson like
polytropic
equation of state with a first order phase transition taking
place at density $\rho_1=3.2\rho_0$ to $\rho_2=4\rho_0$.
\label{rot}  }
\end{figure}

The rotating version of the Tolman-Oppenheimer-Volkoff
equation (\ref{PaGR}) is now solved for a rotating neutron star of mass   
$M=1.4M_\odot$ with the BJP EoS including a first order phase transition. 
In Fig.\ \ref{rot} we show 
the central density, moment of inertia, braking index, star radius and
radius of the interface ($R_0$) as function of the scaled angular
velocity. It is important to note that $R_0\propto
\sqrt{\Omega_0^2-\Omega^2}$ for angular velocities just below the
critical value $\Omega_0$. The qualitative behavior of the
neutron star with the BJP EoS and a first order phase
transition is the same as for our simple analytic example of two
incompressible fluids examined above.  Generally, it is the finite
density difference between the phases that is important and leads to a
term in the moment of inertia proportional to
$(\Omega_0^2-\Omega^2)^{3/2}$ as in Eq.\  (\ref{I}).

The moment of inertia increases with angular velocity. Generally, for a
first order phase transition we find for 
$\Omega \raisebox{-.5ex}{$\stackrel{<}{\sim}$}\Omega_0$
(see also Eq.\  (\ref{I}) and Fig. (\ref{rot}))
\begin{equation}
  I = I_0\left( 1+\frac{1}{2}c_1\frac{\Omega^2}{\Omega_0^2} -\frac{2}{3}c_2 
                (1-\frac{\Omega^2}{\Omega_0^2})^{3/2} + ...
      \right) . 
  \label{Igen}
\end{equation}
For the two incompressible fluids with momentum of inertia given by
Eq.\  (\ref{I}), the small expansion
parameters are $c_1=\omega_0^2$ and 
$c_2=(5/2)\omega_0^3(\rho_2/\rho_1-1)/(2\rho_2/
\rho_1-1-\omega_0^2)^{3/2}$; for $\Omega>\Omega_0$ the $c_2$ term is absent. 
For the BJP we find from Fig.\ (1) that
$c_2\simeq 0.07\simeq 2.2\omega_0^3$. Generally, we find that the
coefficient $c_2$ is proportional to the density difference between the 
two coexisting phases and to the critical angular velocity to the third power,
$c_2\sim (\rho_2/\rho_1-1)\omega_0^3$. The scaled critical angular velocity
$\omega_0$ can at most reach unity for submillisecond pulsars.

To make contact with observation we consider the temporal behavior
of angular velocities of pulsars. The pulsars slow down at a rate
given by the loss of rotational energy which we shall assume is
proportional to the rotational angular velocity to some power
(for dipole radiation $n=3$)
\begin{equation}
  \frac{d}{dt} \left(\frac{1}{2}I\Omega^2\right) = 
                  -C \Omega^{n+1}. 
   \label{dE}
\end{equation}
With the moment of inertia given by Eq. (\ref{Igen})
the angular velocity will then decrease with time as
\begin{eqnarray}
  \frac{\dot{\Omega}}{\Omega} &=& -\frac{C\Omega^{n-1}}{I_0}
  \left( 1-c_1\frac{\Omega^2}{\Omega_0^2}
          -c_2\sqrt{1-\frac{\Omega^2}{\Omega_0^2}} \right) \nonumber\\
  &\simeq& -\frac{1}{(n-1)t}
  \left( 1-c_2 \sqrt{1-(\frac{t_0}{t})^{2/(n-1)}} +....\right), 
  \label{dOmega}
\end{eqnarray}
for $t\ge t_0$. Here,
the time after formation of the pulsar is, using Eq.\  (\ref{dE}),
related to the angular velocity as
$t\simeq t_0(\Omega_0/\Omega)^{n-1}$ and 
$t_0=I_0/((n-1)C\Omega_0^{n-1})$, is the critical time where
a phase transition occurs in the center. For earlier times $t\le t_0$ there
is no dense core and Eq. (\ref{dOmega}) applies when setting $c_2=0$
The critical angular velocity is $\Omega_0=\omega_0\sqrt{2\pi \rho_1}\simeq
6 kHz$ for the BJP EoS, i.e., comparable to a millisecond binary
pulsar. Applying these numbers to, for example, the Crab pulsar we find that
it would have been spinning with critical angular velocity approximately 
a decade after the Crab supernova explosion.

The braking index depends on the second derivative $I''=dI/d^2\Omega$
of the moment of inertia and thus diverges (see Fig. (1))
as $\Omega$ approaches $\Omega_0$ from below
\begin{eqnarray}
     n(\Omega) &\equiv& \frac{\ddot{\Omega}\Omega}{\dot{\Omega}^2} 
    \simeq n - 2c_1\frac{\Omega^2}{\Omega_0^2}
    +c_2\frac{\Omega^4/\Omega_0^4}{\sqrt{1-\Omega^2/\Omega_0^2}} \,.\label{n}
\end{eqnarray}
For $\Omega\ge\Omega_0$ the term with $c_2$ is absent.

We now discuss possible phase transitions in interiors of neutron
stars.  The quark and nuclear matter mixed phase described in
\cite{Glendenning} has continuous pressures and densities. There are
no first order phase transitions but at most two second order phase
transitions. Namely, at a lower density, where quark matter first
appears in nuclear matter, and at a very high density, where all
nucleons are finally dissolved into quark matter. The latter does,
however, not occur if the star is gravitationally unstable at such
high central densities.  In second-order
phase transitions the pressure is a continuous function of density
resulting in a more gentle transition and we find a continuous braking
index.  This mixed phase does, however, not include local surface and
Coulomb energies of the quark and nuclear matter structures. As shown
in \cite{HPS,HH} there can be an appreciable surface and Coulomb
energy associated with forming these structures and if the interface
tension between quark and nuclear matter is too large, the mixed phase
is not favored energetically. The neutron star will then have a core
of pure quark matter with a mantle of quark matter surrounding it and
the two phases are coexisting by a first order phase transition.  For
a small or moderate interface tension the quarks are confined in
droplet, rod- and plate-like structures \cite{HPS,HH} as found in the
inner crust of neutron stars \cite{LPR}. It is found that normal
nuclear matter exists only at moderate densities, $\rho\sim
1-2\rho_0$.  With increasing density droplets of quark matter form in
nuclear matter, then they merge into rod- and later plate-like
structures.  At even higher densities the structures invert forming
plates, rods and droplets of nuclear matter in quark matter.  Finally
pure quark matter is formed at very high densities unless the star
already has exceeded its maximum mass.  Due to the finite Coulomb and
surface energies associated with forming these structures, the
transitions change from second order to first order at each
topological change in structure.  If the surface tension is very small
so are the surface and Coulomb energies of the structures and the
transitions will be only weakly first order.  If a Kaon condensate
appears it may also have such structures \cite{Schaffner}.
Pion condensates \cite{pion}, crystalline nuclear matter \cite{vijay},
hyperonic or magnetized matter, etc. may provide
other first order phase transitions.

There may also be other transitions in neutron stars. The glitches
observed in the Crab, Vela, and a few other pulsars are probably due
to quakes occurring in solid structures such as the crust, 
superfluid vortices or possibly the quark matter lattice in the
core \cite{HH}. These glitches are very small $\Delta\Omega/\Omega\sim
10^{-8}$ and have a characteristic healing time.  In \cite{GPW} a
drastic softening of the equation of state by a phase
transition to quark matter leads to a sudden contraction of the
neutron star at a critical angular velocity and shows up in a
backbending moment of inertia as function of frequency. As a result,
the star will become unstable as it slows down, will suddenly decrease
its moment of inertia and create a large glitch. Similarly, if the
matter undergoes a phase transition at a critical temperature but is
supercooled as the star cools down by neutrino emission, a large
glitch will occur when the matter transforms to its normal state. If
the cooling is continuous the temperature will decrease with star
radius and time and the phase transition boundary will move inwards.
The two phases could, e.g., be quark-gluon/nuclear matter
or a melted/solid phase. In the latter case the size of the hot
(melted) matter in the core is slowly reduced as the temperature drops
freezing the fluid. Melting temperatures have been estimated in
\cite{LPR,melt} for the crust and in \cite{HPS} for the quark matter
mixed phase.  Depending on whether the matter contracts as it freezes
as most terrestial metals or expands as ice, the cooling will separate
the matter in a liquid core of lower or higher density respectively
and a solid mantle around.  When the very core freezes we have a
similar situation as when the star slows down to the critical angular
velocity, i.e., a first order phase transition occurs right at the
center. Consequently, similar behavior of moment of inertia, angular
velocities, braking index may occur as in
Eqs. (\ref{Igen},\ref{dOmega},\ref{n}) replacing $\Omega(t)$ with
$T(t)$.

In summary, we have shown that if a first order phase transitions is
present at central densities of neutron stars, it will show up in
moment of inertia and consequently also in angular velocities in a
characteristic way.
For example, the slow down of the angular velocity
has a characteristic behavior $\dot{\Omega}\sim
c_2\sqrt{1-t/t_0}$ and the braking index diverges as $n(\Omega)\sim
c_2/\sqrt{1-\Omega^2/\Omega_0^2}$ (see Eqs. (\ref{dOmega},\ref{n})).
The magnitude of the signal generally depends on the density
difference between the two phases and the critical angular velocity
$\omega_0=\Omega_0/\sqrt{2\pi G\rho_1}$ such that
$c_2\sim(\rho_2/\rho_1-1)\omega_0^3$.  The observational consequences
depend very much on the critical angular velocity $\Omega_0$, which
depends on the equation of state employed, at which density the phase
transition occurs and the mass of the neutron star. By studying a
range of angular velocities for a sample of different star masses the
chance for encountering a critical angular velocity increases.
Eventually, one may be able to cover the full range of central
densities and find all first order phase transitions up to a certain size
determined by the experimental resolution.  Since the size of the
signal scales with $\Omega_0^3$ the transition may be best observed in
rapidly rotating pulsars such as binary pulsars or pulsars recently
formed in supernova explosion and which are rapidly slowing
down. Carefully monitoring such pulsars may reveal the characteristic
behavior of the angular velocity or braking index as
described above which is a signal of a first order phase transition in
dense matter.

\acknowledgments
We would like to thank Gordon Baym, Larry McLerran, Ben Mottelson and
Chris Pethick for valuable comments.

\end{multicols}

\clearpage

\end{document}